\documentclass[aps,prb,reprint,superscriptaddress]{revtex4-2}
\usepackage{graphicx}
\usepackage{amsmath}
\usepackage{amssymb}
\usepackage{natbib}

\begin{document}

\title{Two-dimensional ferroelectricity induced by octahedral rotation distortion in perovskite oxides}
\author{Ying Zhou}
\affiliation{School of Materials Science and Physics, China University of Mining and Technology, Xuzhou 221116, China}
\author{Shuai Dong}
\affiliation{School of Physics, Southeast University, Nanjing 211189, China}
\author{Changxun Shan}
\author{Ke Ji}
\author{Junting Zhang}
\email{juntingzhang@cumt.edu.cn}
\affiliation{School of Materials Science and Physics, China University of Mining and Technology, Xuzhou 221116, China}

\begin{abstract}
Two-dimensional (2D) ferroelectricity has attracted extensive attention since its discovery in the monolayers of van der Waals materials. Here we show that 2D ferroelectricity induced by octahedral rotation distortion is widely present in the perovskite bilayer system through first-principles calculations. Perovskite tolerance factor plays a crucial role in the lattice dynamics and ground-state structure of the perovskite monolayers and bilayers, thus providing an important indicator for screening this hybrid improper ferroelectricity. Generally, the ferroelectric switching via an orthorhombic twin state has the lowest energy barrier. Epitaxial strain can effectively tune the ferroelectric polarization and ferroelectric switching by changing the amplitude of octahedral rotation and tilt distortion. The increasing compressive strain causes a polar to non-polar phase transition by suppressing the tilt distortion. The cooperative effect of octahedral distortion at the interface with the substrate can reduce the energy barrier of reversing rotation mode, and even change the lowest-energy ferroelectric switching path.
\end{abstract}
\maketitle

\section{Introduction}
The two-dimensionalization of ferroelectricity has attracted considerable interest in recent years due to its potential applications in high-density nonvolatile memory, low energy consumption and miniaturization of electronic devices \cite{Dawber2005,Scott2007,Fei2016,Ding2017,Osada2019}. However, the existence of two-dimensional (2D) ferroelectricity has been a long open question \cite{Dawber2005}. Conventional ferroelectric films, especially those with out-of-plane ferroelectricity desired in devices, generally exhibit critical thickness effect due to the depolarization fields or surface and interface effects \cite{Woo2007,Junquera2003,Fong2004}. For example, ferroelectricity disappears when some perovskite ferroelectric films are reduced to the thickness of a few unit cells \cite{Junquera2003,Fong2004}. However, some theoretical and experimental studies have shown the possible absence of critical thickness, in which the multi-domain states, and the interface effect with a conducting electrode or insulating substrate have been found to be crucial for neutralizing the depolarization field and stabilizing the ferroelectricity in ultrathin films \cite{Kolpak2005,Gao2017,Baker2020}. Very recently, 2D ferroelectricity down to the monolayer limit has been demonstrated in van der Waals materials \cite{Chang2016,Liu2016,Zhou2017}. Some novel ferroelectric mechanisms and phenomena have emerged in these 2D materials, such as switchable spontaneous polarization in elemental monolayers and 2D metals \cite{Li2017,Xiao2018,Fei2018}.

It is of great significance to achieve 2D ferroelectricity beyond van der Waals materials, since conventional ferroelectrics are mostly concentrated in transition-metal oxides, especially perovskite oxides \cite{Dawber2005,Scott2007}. Recently, the successful growth of the freestanding perovskite oxides down to the monolayer limit has paved the way for the design of some functional properties based on 2D perovskite oxides \cite{Ji2019,Hong2017}. This also provides an opportunity to resolve some controversial issues involving 2D ferroelectricity, such as whether the conventional ferroelectric mechanisms can be maintained to the 2D limit \cite{Gao2017}, or whether a novel ferroelectric mechanism emerges in the 2D perovskite oxides.

In perovskite, the proper ferroelectricity is usually derived from an electronic mechanism, the second-order Jahn-Teller effect, which is commonly present in perovskite oxides with $d^0$-configuration transition-metal ions at the \emph{B} site \cite{Bersuker2013}. The hybrid improper ferroelectricity, \emph{i.e}., octahedral rotation-induced ferroelectricity, originates from structural geometry effects and thus has no restriction on the electronic configuration of \emph{B}-site ions \cite{Benedek2011,Benedek2012}. This improper ferroelectricity has been demonstrated in layered perovskites and perovskite superlattice \cite{Bousquet2008,Akamatsu2014,Mulder2013,Oh2015,Yoshida2018,Yoshida2018a}, but is absent in perovskite bulks due to the preservation of inversion symmetry by octahedral rotation. Until recently, theoretical studies suggested that octahedral-rotation induced ferroelectricity can occur in 2D perovskites \cite{Lu2017,Zhang2020,Zhou2021}. However, there are still some remaining issues to be resolved, such as the key influencing factors and the universality of this ferroelectricity.

In this paper, a series of perovskite monolayers and bilayers with formula $A_2B{\rm O}_4$ and $A_3B_2{\rm O}_7$ (\emph{A}=Ca, Sr; \emph{B}= Ti, Zr, Si, Ge, Sn), respectively, are selected to study the 2D ferroelectricity based on perovskite oxides. The divalent alkaline-earth metal ions are selected at \emph{A} site to form a non-polar surface, and the tetravalent transition-metal ions with $d^0$ configuration (Ti, Zr) and group-\uppercase\expandafter{\romannumeral4} ions (Si, Ge, Sn) are selected at \emph{B} site for comparison. We study their lattice dynamics, structure, ferroelectricity, as well as strain and interface effects by using first-principles calculations. We reveal the correlation between lattice dynamics and ground-state structure, and the role of perovskite tolerance factor. We show that the proper ferroelectricity is absent in these 2D perovskite oxides, while the hybrid improper ferroelectricity is widely present in the perovskite bilayers with small tolerance factors and can be significantly tuned by epitaxial strain. The interface effect can also have a significant impact on the ferroelectric switching.
\section{Computational details}
The first-principles calculations based on density functional theory (DFT) were performed using the projector-augmented wave (PAW) method \cite{Bloechl1994}, as implemented in the Vienna \emph{ab initio} simulation package (VASP6.1) \cite{Kresse1996}. The Ca $3s3p4s$, Sr $4s4p5s$, Ti $3p3d4s$, Zr $4p4d5s$, Si $3s3p$, Ge $4s4p$, Sn $5s5p$, and O $2s2p$ electrons were treated as valence electrons in the PAW potentials. The Perdew-Burke-Ernzerhof functional modified for solids (PBEsol) \cite{Perdew2008} was used as the exchange-correlation functional. The $\sqrt{2}\times\sqrt{2}\times1$ and $\sqrt{2}\times\sqrt{2}\times2$ perovskite supercells were used to construct distorted structures caused by various octahedral rotation for the perovskite monolayers and bilayers, respectively. A vacuum space of 20 {\AA} was imposed in order to avoid interaction between neighboring periodic images. We used a plane-wave cutoff energy of 600 eV for the plane-wave expansion, and a $\varGamma$-centered $7\times7\times1$ \emph{k}-point mesh for the Brillouin zone integration. The in-plane lattice constants and internal atomic coordinates of each structural phase were relaxed until the Hellman-Feynman force on each atom is less than 0.01 eV/{\AA}. A convergence threshold of $10^{-6}$ eV was used for the electronic self-consistency loop. Phonon band structures were calculated using density functional perturbation theory (DFPT) \cite{Gonze1997}. The phonon frequencies and corresponding eigenmodes were calculated on the basis of the extracted force-constant matrices, as implemented in the PHONOPY code \cite{Togo2015}. The non-analytical term correction was used to calculate the LO-TO splitting of phonon frequency \cite{Gonze1997}. In order to confirm the thermodynamic stability of the ground-state phases, the \emph{ab} initio molecular dynamics simulations at 300 K were performed in a canonical ensemble, and a large supercell containing 216 atoms was used to minimize the constraint of periodic boundary conditions. The Born effective charge was calculated using DFPT method, and the ferroelectric polarization was calculated based on the Born effective charge obtained. The climbing nudged elastic band method \cite{Sheppard2012} was used to determine the energy barrier of different ferroelectric switching paths. The ISOTROPY tool \cite{Campbell2006} was used to aid with the group-theoretical analysis.

\section{results and discussion}

\subsection{Lattice dynamics}
\begin{figure}
\centering
\includegraphics*[width=0.45\textwidth]{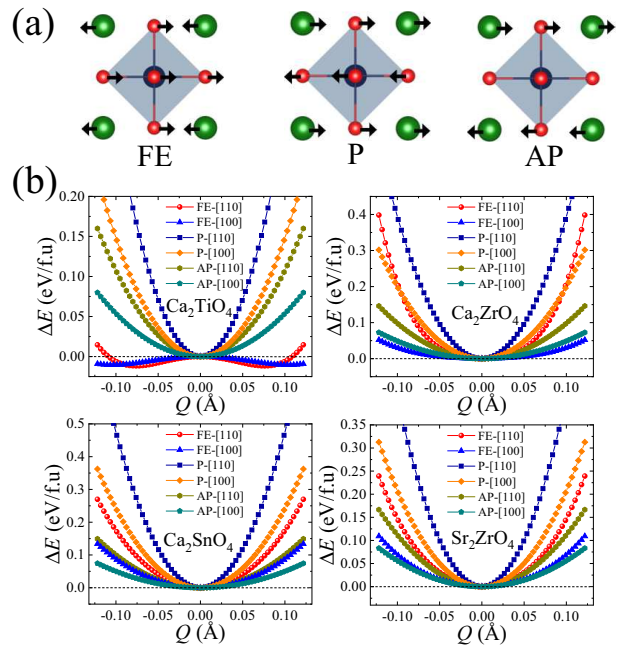}
\caption{\label{Fig1} (a) Ferroelectric (FE), polar (P), and antipolar (AP) soft modes appearing in perovskite monolayers. The arrow represents the direction of ion displacement. (b) The energy of the distorted structure caused by each soft mode as a function of the mode amplitude. The ion displacement of the soft mode is along the [100] and [110] directions, respectively.}
\end{figure}
We first studied the lattice dynamics of the selected 2D perovskite oxides and revealed its correlation with perovskite tolerance factor. The phonon spectra of the prototype phases (space group $P4/mmm$) of the perovskite monolayers are shown in Fig. S1 of the Supplemental Material \cite{SM}. Only the prototype phase of the Sr$_2$SiO$_4$ monolayer with a maximum tolerance factor ($t=1.127$) exhibits dynamic stability. The remaining monolayers all appear unstable modes at the high symmetry \emph{M} point of the Brillouin zone, where the soft mode with the lowest frequency represents the octahedral rotation distortion. Note that the octahedral rotation distortion can be divided into rotation and tilt modes in 2D perovskites, which represent the octahedral rotation around the out-of-plane and in-plane axes, respectively. The lowest-frequency soft mode belongs to the rotation mode, while the tilt mode is stable for all monolayers. The rest of the soft modes involves the deformation of the octahedron. The monolayers with $t >1$ and $t <1$ respectively exhibit single and multiple soft modes at \emph{M} point, expect for the Ca$_2$GeO$_4$ monolayer with $t=1.012$ (see Table \uppercase\expandafter{\romannumeral1}). In addition, the monolayers with $t < 1$ (except Sr$_2$SnO$_4$) also show some soft modes in the center of the Brillouin zone, involving ferroelectric, polar and antipolar modes, as shown in Fig. 1(a). For the polar and antipolar modes, the ion displacements in the upper and lower surfaces are the same and opposite respectively. Interestingly, the ferroelectric soft mode appears in the Ca$_2$SnO$_4$ monolayer without transition-metal ion. Therefore, the ferroelectric soft mode may not be driven by the electronic mechanism related to the second-order Jahn-Teller effect \cite{Bersuker2013}, but rather related to structural geometric effect, namely cation size mismatch.

Then we calculated the energy gain caused by the freezing of each soft mode. Among the soft modes appearing at \emph{M} point, only the rotate mode can reduce the energy of the system (see Fig. S2 \cite{SM}). In addition, although the tilt mode is stable, its freezing can also cause a slightly reduction in energy. For all the soft modes appearing at $\varGamma$ point, only the ferroelectric mode of the Ca$_2$TiO$_4$ monolayer shows a typical double-well type energy curve, while the remaining soft modes show a parabolic energy curve, as shown in Fig.1 (b). Therefore, not all soft modes can generate energy gain, which may be related to the fact that the anharmonic effect is not considered in the calculation of phonon band structures. Similar phenomena have also been found in other perovskite and layered perovskite oxides \cite{Cammarata2015,Zhang2020,Zhou2021}.

\begin{table}
\centering
\includegraphics*[width=0.48\textwidth]{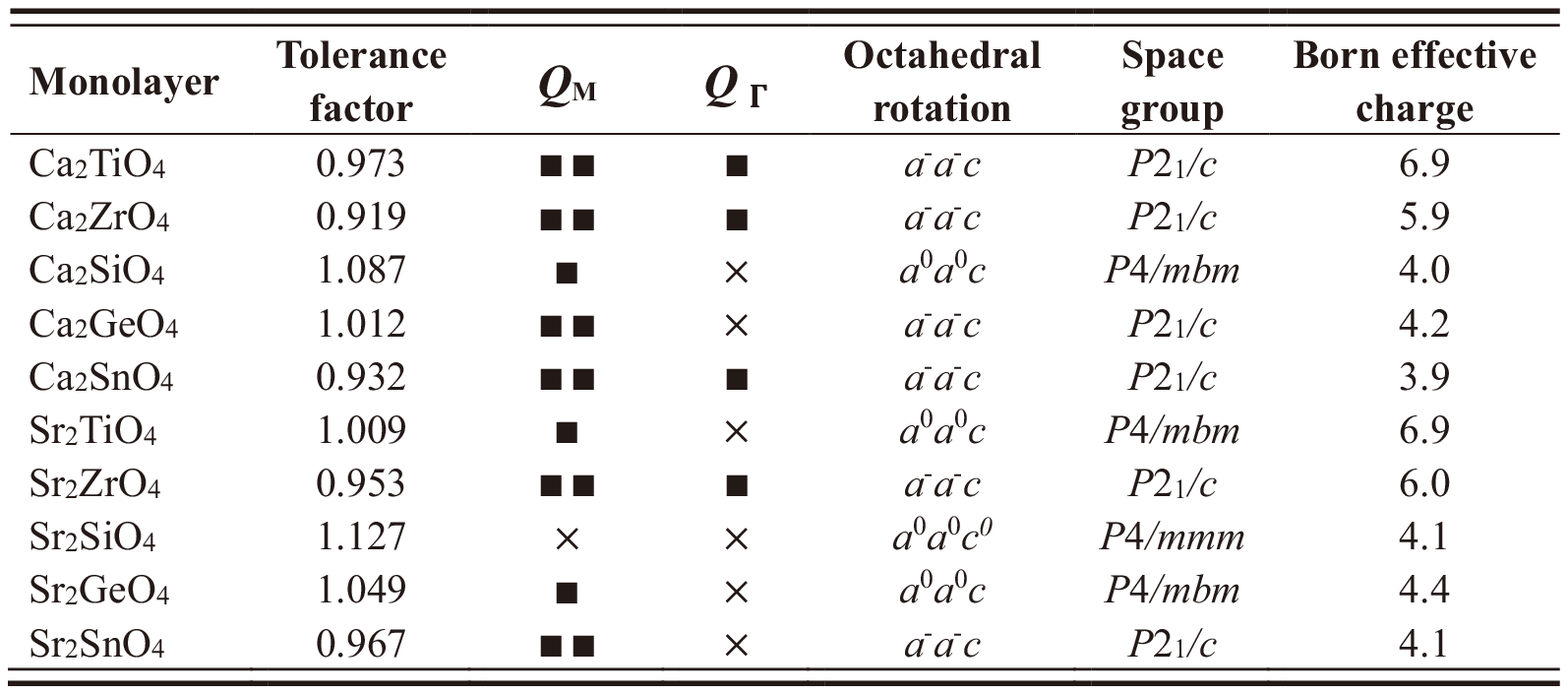}
\caption{\label{table 1} Tolerance factor, soft modes of prototype phase, octahedral rotation and space group of ground-state structure, and Born effective charge of \emph{B}-site ions of the peprovskite monolayers. The symbols "$\scriptstyle{\blacksquare}$" and "$\times$" respectively represent the presence and absence of the soft mode at the corresponding high symmetry points, and "$\scriptstyle{\blacksquare\blacksquare}$" represents the presence of multiple soft modes at \emph{M} point.}
\end{table}
For the perovskite bilayers, only the prototype phase of the Sr$_3$Si$_2$O$_7$ bilayer exhibits dynamic stability (see Fig. S3 \cite{SM}), similar to the monolayer system. The remaining bilayers show soft modes at \emph{M} point, and the bilayers with relatively small tolerance factor show multiple soft modes. The lowest-frequency soft modes involve almost degenerate in-phase (IR) and out-of-phase (OR) octahedral rotation modes. The bilayers with $t < 1$ (except Sr$_3$Sn$_2$O$_7$) show multiple soft modes at $\varGamma$ point, involving ferroelectric, polar and antipolar modes [see Fig. 2(a)], similar to their monolayers. The structural distortion caused by the rotation or tilt mode can reduce the energy of the bilayer (see Fig. S2 \cite{SM}), similar to the monolayer system. However, different from the monolayer system, the $Q_\varGamma$ soft modes of the bilayers generally show double-well type energy curve, as shown in Fig. 2(b). Their maximum energy gain is inversely proportional to the tolerance factor, for example, the largest energy gain occurs in the Ca$_3$Zr$_2$O$_7$ bilayer with the smallest tolerance factor.

\begin{figure}
\centering
\includegraphics*[width=0.45\textwidth]{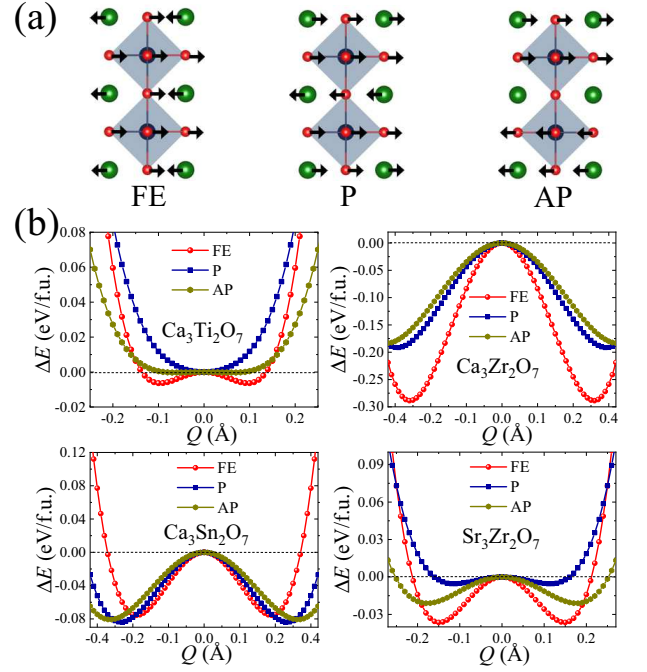}
\caption{\label{Fig1} (a) Ferroelectric (FE), polar (P), and antipolar (AP) soft modes appearing in peroovskite bilayers. (b) The energy of the distorted structure caused by each soft mode as a function of the mode amplitude.}
\end{figure}

\begin{table}
\centering
\includegraphics*[width=0.48\textwidth]{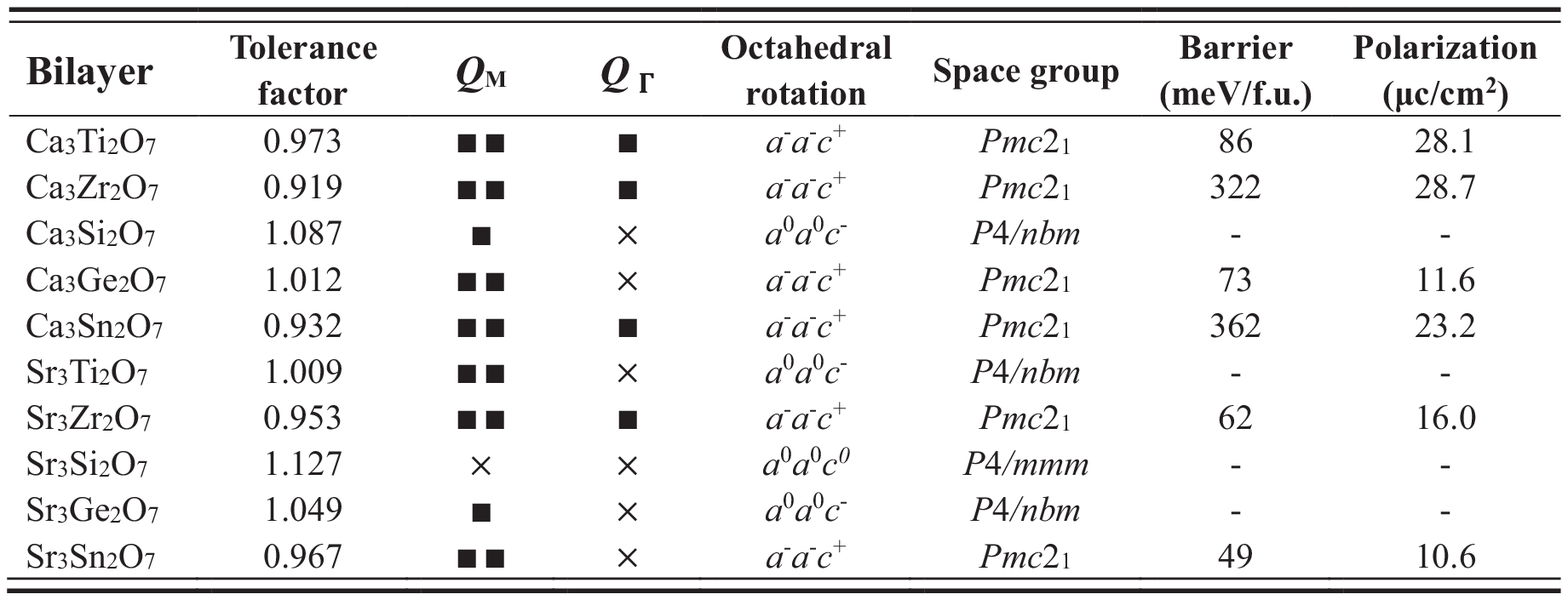}
\caption{\label{table 1} Tolerance factor, soft modes of prototype phase, octahedral rotation and space group of ground-state structure, the lowest energy barrier of ferroelectric switching, and ferroelectric polarization of the peprovskite bilayers. The symbol "-" represents the absence of ferroelectricity in the ground-state phase.}
\end{table}

\subsection{Ground-state structure}
Next, in order to determine the ground-state structure, we considered the possible coupling between different structural distortion modes. For the perovskite monolayer, in the presence of octahedral rotation ($Q_{M_2^+}$) or tilt ($Q_{M_5^+}$) mode, the ferroelectric ($Q_{\varGamma_5^-}$), polar ($Q_{\varGamma_5^-}$) and antipolar ($Q_{\varGamma_5^+}$) modes will not reduce the energy of the system. In addition, in the distorted structure established by $Q_{M_2^+}$ (or $Q_{M_5^+}$) $\oplus$ $Q_{\varGamma_5^-}$ (or $Q_{\varGamma_5^+}$) modes, the latter basically disappears after structural optimization. However, in the distorted structure caused by $Q_{M_2^+}\oplus Q_{M_5^+}$ modes, the antipolar $Q_{\varGamma_5^+}$ mode can coexist with them. This can be attributed to the trilinear coupling $Q_{M_2^+}Q_{M_5^+}Q_{\varGamma_5^+}$ allowed by symmetry.

Then we analyzed the structural symmetry of various octahedral rotation types, and calculated the energy of each structural phase (see Tables S1 and S2 \cite{SM}) by optimizing the atomic coordinates and the in-plane lattice constants under fixed symmetry. As shown in Table \uppercase\expandafter{\romannumeral1}, only the ground-state structure of the Sr$_2$SiO$_4$ monolayer maintains the prototype phase, consistent with the results of lattice dynamics. The ground-state structure of the remaining monolayers are divided into two categories, basically with the tolerance factor $t=1$ as the boundary, that is, when the tolerance factor is greater than and less than 1, the ground-state structure becomes the $P4/mbm$ ($a^0a^0c$ type octahedral rotation) and $P2_1/c$ ($a^-a^-c$ type) phases, respectively. The only exception is the Ca$_2$GeO$_4$ monolayer, which has a tolerance factor ($t=1.012$) slightly greater than 1, but shows $a^-a^-c$ type octahedral rotation. This exception is consistent with the results of its lattice dynamics. When the prototype phase shows single (\emph{i.e.}, only rotate mode) and multiple soft modes at \emph{M} points, the ground-state phase exhibits only rotation ($a^0a^0c$) and rotation plus tilt ($a^-a^-c$) distortion, respectively.

For the perovskite bilayers, only the ground-state structure of the Sr$_3$Si$_2$O$_7$ bilayer does not show any structural distortion (see Table 2), similar to its monolayer. The ground-state structure of the bilayers with $t>1$ (except Ca$_3$Ge$_2$O$_7$) exhibits the $P4/nbm$ phase caused by a single OR ($Q_{M_4^-}$) mode, while the ground-state structure of the bilayers with $t<1$ is the polar $Pmc2_1$ phase established by IR ($Q_{M_2^+}$) plus tilt ($Q_{M_5^-}$) distortion. The octahedral rotation type ($a^-a^-c^+$) of these polar bilayers is not reconstructed with respect to their perovskite bulk and layered bulk phases \cite{Mulder2013,Oh2015,Yoshida2018,Yoshida2018a}. In this polar ground-state phase, the polar mode ($Q_{\varGamma_5^-}$) [see Fig. 2(a)] emerges and is coupled with the IR and tilt modes, which is a manifestation of the trilinear coupling $Q_{M_2^+}Q_{M_5^-}Q_{\varGamma_5^-}$. This polar mode is also essential for maintaining the octahedral rotation distortion of this ground-state phase (see Fig. S4 \cite{SM}). There is another trilinear coupling for the perovskite bilayers, $Q_{M_4^-}Q_{M_5^-}Q_{\varGamma_5^+}$. This trilinear coupling is responsible for the hybrid improper antiferroelectricity, which has been demonstrated in layered perovskites with Ruddlesden-Popper phase \cite{Yoshida2018}.

\subsection{Ferroelectricity}
The transition-metal ions with $d^0$ configuration (Ti$^{4+}$ and Zr$^{4+}$) in the 2D perovskites show abnormally large Born effective charge (see Table. \uppercase\expandafter{\romannumeral1}), similar to the perovskite bulks \cite{Ghosez1999}, indicating the possible existence of the proper ferroelectricity. In fact, this proper ferroelectricity related to the \emph{B}-site ion with empty \emph{d} orbital has been predicted to exist in 2D perovskites \cite{Lu2017}. However, the structural phase caused by ferroelectric or polar mode individually does not become the ground-state phase of these selected perovskite monolayers and bilayers (see Tables S1 and S2 \cite{SM}). In the perovskite monolayer, each octahedral distortion mode retains the inversion center at the center of the octahedron, which rules out the possibility of ferroelectricity caused by octahedral distortion. When extended to the perovskite bilayer, the combination of two octahedral distortion modes with different inversion centers, such as IR plus tilt modes, can break the inversion symmetry, leading to the appearance of polar phase.

\begin{figure}
\centering
\includegraphics*[width=0.4\textwidth]{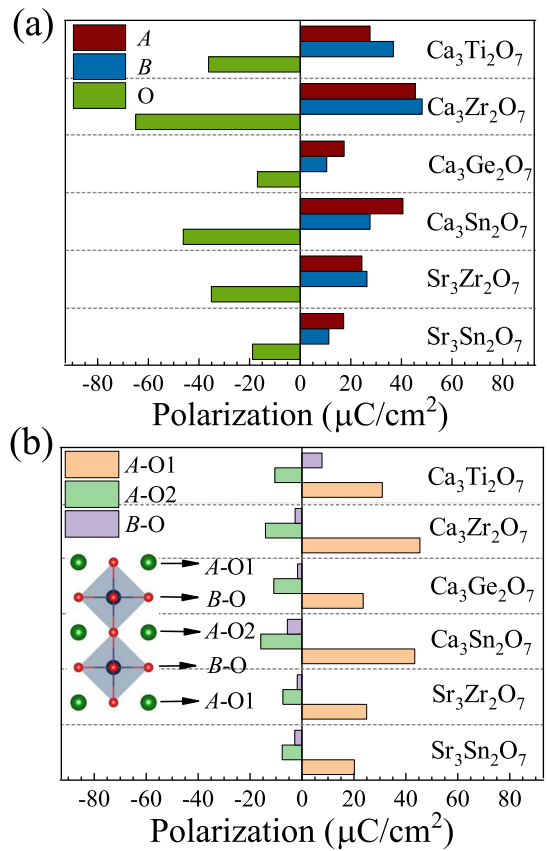}
\caption{\label{Fig3} (a) Ion-resolved and (b) layer-resolved polarization of the polar perovskite bilayers.}
\end{figure}

We next focused on the polar ground-state phase of the perovskite bilayers. The phonon spectrum calculation and the first-principles molecular dynamics simulation respectively confirmed its dynamic and thermodynamic stability (see Fig. S5 \cite{SM}). Although the IR plus tilt modes can directly establish a polar phase in the absence of polar mode, the polar distortion
is responsible for the appearance of ferroelectric polarization. The ferroelectric polarization disappears if we remove the polar distortion from the ground-state structure and retain the octahedral rotation distortion. In contrast, if we remove the octahedral rotation distortion and keep the polar distortion unchanged, the calculated polarization does not change, confirming that the ferroelectric polarization originates entirely from the polar distortion.

Figure 3 shows the contribution of different ions and atomic layers to the ferroelectric polarization. The polar displacements of cations and anions induces the opposite polarization, and the polarization induced by the \emph{A} and \emph{B} cations is close. This indicates that the polarization originates from the polar mode rather than the ferroelectric mode shown in Fig. 2(a). In general, the perovskite bilayers with smaller tolerance factor have larger polarization induced by \emph{A} and O ions. The polarization induced by the two surface \emph{A}-O layers is opposite to and much larger than that of the intermediate \emph{A}-O layer, and the polarization induced by the \emph{B}-O layers is rather weak, as shown in Fig. 3(b), indicating that the ferroelectric polarization is mainly derived from the ion displacement of the surface layer. Note that unlike other polar bilayers, the polarization induced by the \emph{B}-O layers in the Ca$_3$Ti$_2$O$_7$ bilayer is considerable and in the same direction as that of the \emph{A}-O layer, which may be the reason for its maximum polarization among these bilayers. The magnitude of total ferroelectric polarization does not show an obvious dependence on the tolerance factor (see Table \uppercase\expandafter{\romannumeral2}). For example, the polarization values of the Ca$_3$Ti$_2$O$_7$ and Ca$_3$Zr$_2$O$_7$ bilayers are very close despite the large differences of tolerance factor. The calculated polarization is greater than that of the corresponding layered perovskite bulk phases \cite{Mulder2013,Oh2015,Yoshida2018,Yoshida2018a}.

The reversal of polarization can be caused by switching the senses of rotation of IR or tilt mode individually. Both modes can be reversed by a one-step (OS) or multi-steps (MS) switching \cite{Zhang2020,Zhou2021}. The reversal of IR mode via MS refers to the change of the octahedral rotation direction in the upper and lower layers sequentially, so it undergoes a nonpolar phase established by OR plus tilt modes. For the MS switching of reversing tilt mode, the tilt axis undergoes a 180$^\circ$ rotation in the \emph{ab} plane, so that it passes through an orthogonal twin state with tilt axis perpendicular to its initial direction. Previous experiments have shown that the low switching barrier of the bulk ferroelectric Ca$_3$Ti$_2$O$_7$ is related to the appearance of orthogonal twin domains \cite{Oh2015}. We calculated the energy barrier of different switching paths for the perovskite bilayers with polar ground-state phase (see Fig. 4). The results show that the MS switching path always has a lower energy barrier than the corresponding OS switching path. The MS switching of reversing tilt mode generally has the lowest energy barrier, which may be attributed to the fact that this switching path always maintains IR plus tilt distortion. However, for the Ca$_3$Zr$_2$O$_7$ and Ca$_3$Sn$_2$O$_7$ bilayers with smaller tolerance factors, the MS switching path of reversing IR mode has close or even degenerate energy barrier with respect to this lowest-energy switching path.

\begin{figure}
\centering
\includegraphics*[width=0.48\textwidth]{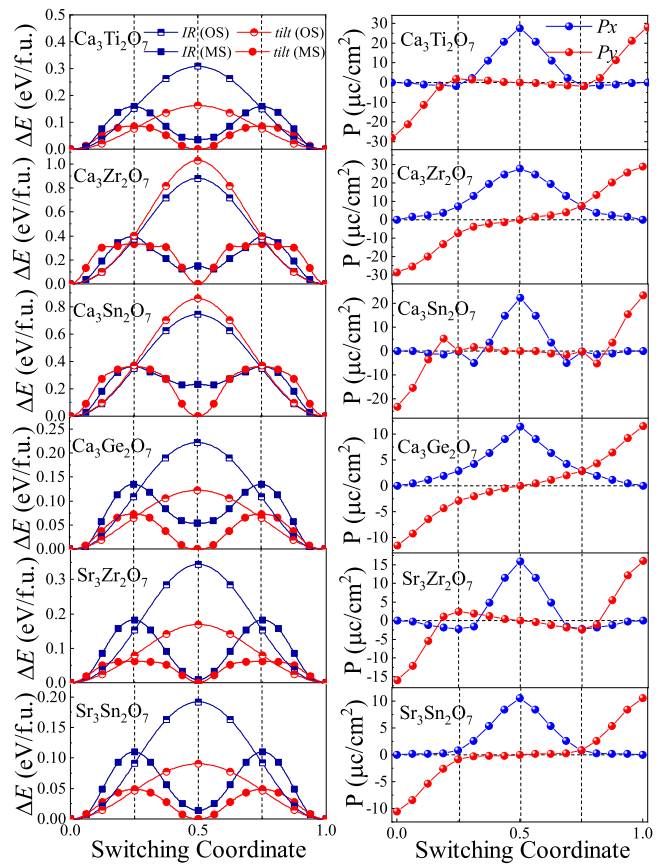}
\caption{\label{Fig3} Total energy of each polar perovskite bilayer as a function of switching coordinate along different ferroelectric switching paths, involving reversing the IR and tilt modes via one step (OS) or multi steps (MS). The panels on the right show the variation of the polarization component along the lowest-energy ferroelectric switching path.}
\end{figure}

The lowest energy barrier is proportional to the amplitude of tilt distortion. For example, the Ca$_3$Zr$_2$O$_7$ and Ca$_3$Sn$_2$O$_7$ bilayers have relatively large amplitude of tilt distortion, resulting in very high energy barrier (see Table \uppercase\expandafter{\romannumeral2}). Therefore, their electric polarization is difficult to be reversed by the electric field. Other polar bilayers have relatively small energy barrier ($<100$ meV/f.u.), of which the Sr$_3$Sn$_2$O$_7$ bilayer has the lowest energy barrier (49 meV/f.u.). The ferroelectricity of some of their corresponding layered bulk phases including Ca$_3$Ti$_2$O$_7$ \cite{Oh2015}, Sr$_3$Sn$_2$O$_7$ \cite{Yoshida2018a}, and Sr$_3$Zr$_2$O$_7$ \cite{Yoshida2018} have been demonstrated experimentally, and the ferroelectric switching barrier of the Ca$_3$Ti$_2$O$_7$ bilayer is very close to the calculated value (82 meV/f.u.) of its layered bulk phase \cite{Nowadnick2016}.

The variation of polarization component during the lowest-energy ferroelectric switching is shown in Fig. 4. The direction of polarization changes continuously in the \emph{ab} plane during the ferroelectric switching. In addition, the magnitude of polarization undergoes a significant change, that is, it is greatly reduced when approaching the barrier phase. This change in polarization stems from the variation of tilt and polar modes during ferroelectric switching, which exhibit similar change trend (see Fig. S6 \cite{SM}). The above results indicate that the octahedral rotation-induced ferroelectricity can be maintained in the perovskite bilayer. This hybrid improper ferroelectricity is likely to be widespread in this 2D system since it is derived from purely geometric effect rather than electronic mechanism. The magnitudes of ferroelectric polarization and the ferroelectric switching barrier can be optimized by selecting cation combinations.

\subsection{Strain and interface effect}
It is known that the octahedral rotation can be significantly affected by the epitaxial strain in the epitaxially grown perovskite films \cite{Lu2016,May2010}. Then we studied the strain effect of the perovskite bilayers grown on a square substrate. The in-plane lattice parameters are fixed to be the same as those of the substrate due to epitaxial matching. For the Ca$_3$Zr$_2$O$_7$ and Ca$_3$Sn$_2$O$_7$ bilayers with smaller tolerance factors, the polar $Pmc2_1$ phase is always the ground-state phase in the entire strain range, that is, the epitaxial strain does not change their octahedral rotation type, as shown in Fig. 5. However, for the remaining four polar bilayers with relatively large tolerance factors, the ground-state structure gradually changes from the polar $Pmc2_1$ phase to the non-polar $P4/nbm$ phase with increasing compressive strain, which can be attributed to the suppression of tilt distortion by compressive strain (see Fig. S7 \cite{SM}). This polar to nonpolar transition has been demonstrated theoretically in some layered $A_3B_2$O$_7$ type perovskite oxides \cite{Lu2016}. The energy curves of these polar bilayers show similar change trend with strain. The energy difference between different structural phases (except the two phases with only tilt distortion) gradually decreases with the increase of compressive strain. The polar ground-state phase is absent in the strained bilayers with larger tolerance factor, that is, the epitaxial strain fails to induce the polar phase in the non-polar bilayers. The ground-state phase ($P4/mmm$) of the strained Sr$_3$Si$_2$O$_7$ bilayer remains the same as its unstrained state. However, for the remaining three non-polar bilayers, the ground-state structure changes from $P4/nbm$ ($a^0a^0c^-$ type octahedral rotation) to $Pmma$ phase ($a^-a^-c^0$) with the increase of tensile strain. This transition is related to the suppression of rotation distortion and the enhancement of tilt distortion by the increasing tensile strain. Similar strain effects on octahedral rotation have been found in perovskite oxide heterostructures \cite{Rondinelli2012a,Choquette2016}.
\begin{figure}
\centering
\includegraphics*[width=0.48\textwidth]{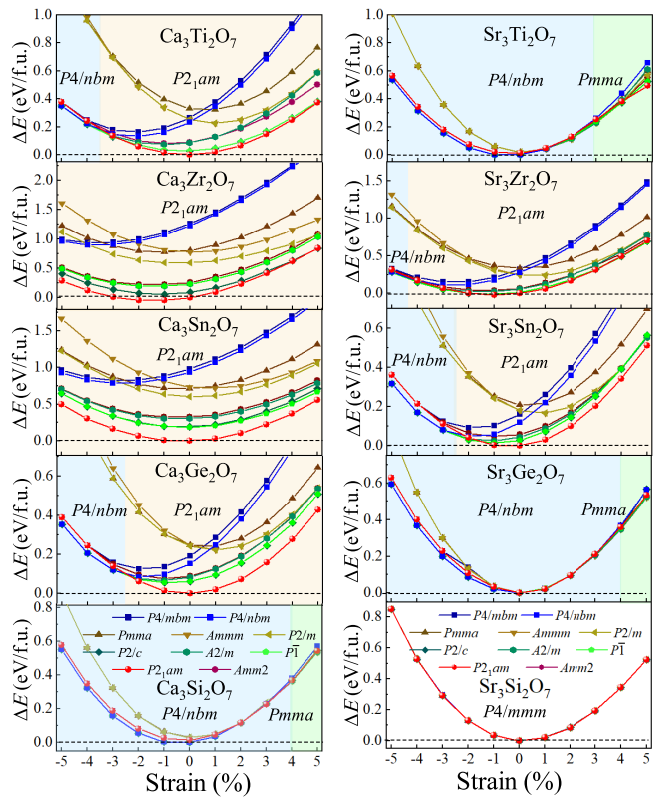}
\caption{\label{Fig4} Change in energy of all structural phases with epitaxial strain. The energy curves of the prototype phase and the structural phases caused by the $Q_\varGamma$ mode alone are not shown due to their high energy. The strain value is defined as $\varepsilon = (d-d_0)/d_0$, where \emph{d} and $d_0$ are the in-plane lattice constants of the strained and unstrained (the prototype phase) bilayers, respectively. The ground-state phases are marked and their ranges are color-coded.}
\end{figure}

The significant variation of the amplitude of the octahedral rotation distortion with strain (see Fig. S7 \cite{SM}) indicates that the epitaxial strain can be used as an effective means to tailor the octahedral rotation-induced ferroelectricity. In the case of the most commonly used SrTiO$_3$ substrate, the Ca$_3$Ti$_2$O$_7$ and Ca$_3$Sn$_2$O$_7$ bilayers grown on this substrate are subjected to tensile ($+2.6\%$) and compressive ($-2.2\%$) strain, respectively. For the Ca$_3$Ti$_2$O$_7$ bilayer, this tensile strain lowers and raises the energy barrier of reversing rotation and tilt modes, respectively. This causes the MS switching of reversing IR mode to become the lowest-energy ferroelectric switching path [see Fig. 6(a)]. In contrast, the compressive strain reduces the energy barrier of reversing tilt mode for the Ca$_3$Sn$_2$O$_7$ bilayer, resulting in the lowest energy barrier for the MS switching of reversing tilt mode (see Fig. S8 \cite{SM}). Its ferroelectric polarization is also correspondingly reduced due to the attenuation of the amplitude of the tilt mode.

\begin{figure}
\centering
\includegraphics*[width=0.45\textwidth]{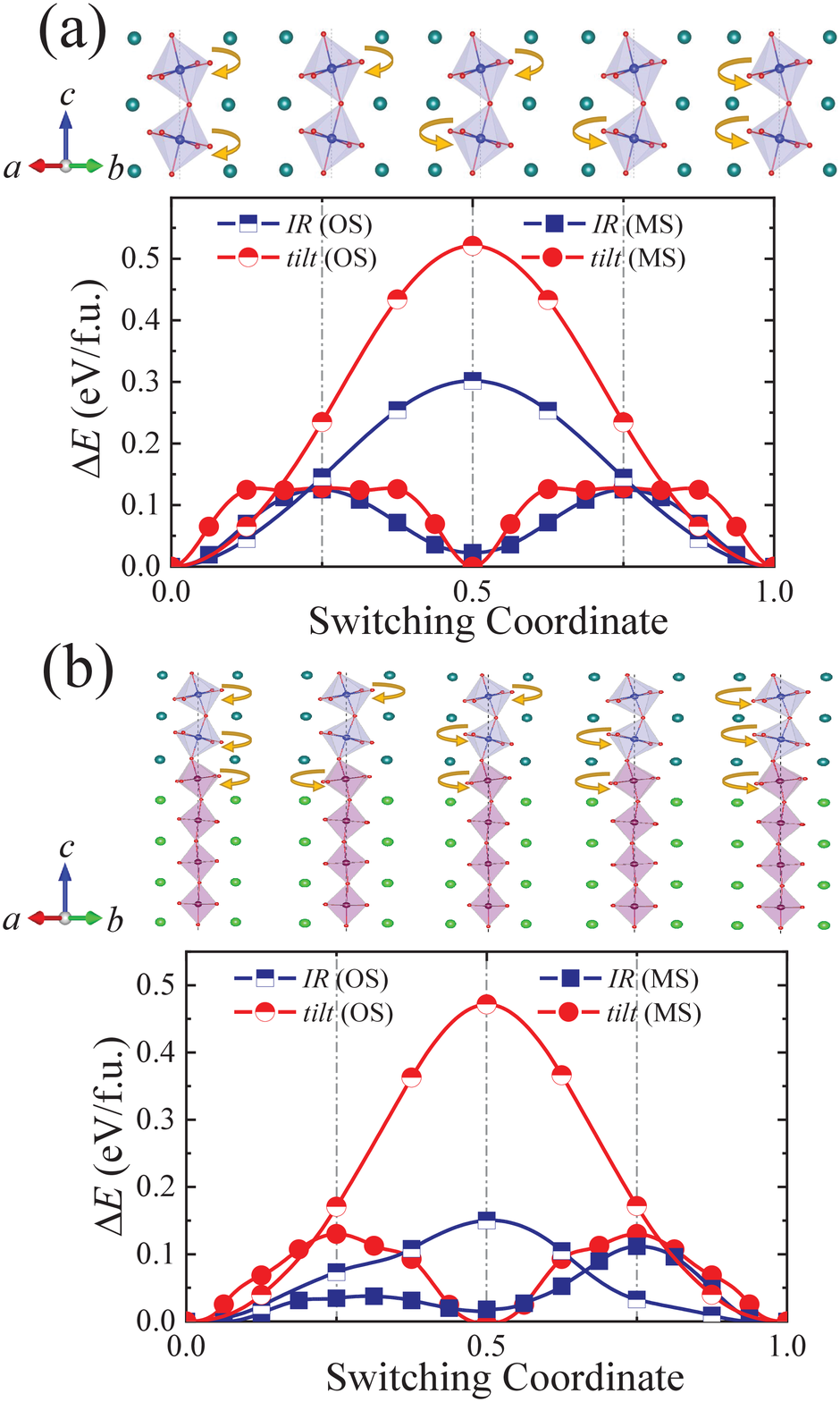}
\caption{\label{Fig5} (a) Change in energy along different ferroelectric switching paths for the tensile strained Ca$_3$Ti$_2$O$_7$ bilayer, whose in-plane lattice constants are fixed to that of the SrTiO$_3$ substrate. The upper subgraph illustrates the MS path of reversing the IR mode. The curved arrow represents the rotation direction of the octahedron around the \emph{c} axis. (b) Change in energy along different ferroelectric switching paths for the heterostructure Ca$_3$Ti$_2$O$_7$/(SrTiO$_3)_4$, and the schematic of the MS path of reversing the IR mode.}
\end{figure}
Next, we used a heterostructure model Ca$_3$Ti$_2$O$_7$/(SrTiO$_3)_4$ to reveal the effect of the interface formed with the substrate. The SrO layer at the bottom of the SrTiO$_3$ substrate was fixed during structural optimization to simulate a thick substrate \cite{An2017,Ji2019a}. The results of electrostatic potential show that the out-of-plane electric field caused by the asymmetry of the heterostructure model is negligible (see Fig. S9 \cite{SM}). In addition, the introduction of dipole correction \cite{Bengtsson1999} does not change the distribution of electrostatic potential and the energy barrier of ferroelectric switching. The structural distortion of the Ca$_3$Ti$_2$O$_7$ bilayer induces octahedral rotation and tilt distortion in the substrate, which is mainly concentrated in the interface SrTiO$_3$ layer [see Fig. 6(b)]. The interface effect does not change the octahedral rotation type and electronic properties of the Ca$_3$Ti$_2$O$_7$ bilayer. However, it has a significant impact on ferroelectric switching. In the heterostructure model, the energy barrier of ferroelectric switching by reversing IR mode (especially OS switching) is significantly reduced, while that by reversing tilt mode is only slightly changed [see Figs. 6(a) and 6(b)]. This is related to the cooperative effect of the octahedral rotation distortion at the interface. The reversal of the IR mode in the bilayer also leads to the reversal of the rotation in the interface SrTiO$_3$ layer. The presence of octahedral rotation distortion in the interface layer resulting in asymmetry in the first half and the last half of the MS switching path of reversing IR mode, which is responsible for the significant difference in the energy of the two barrier phases. A similar interface effect is also observed in the Ca$_3$Sn$_2$O$_7$/(SrTiO$_3)_4$ heterostructure [see Fig. S8(b)] \cite{SM}.

\section{conclusion}
In conclusion, we have systematically studied the lattice dynamics, structure, ferroelectricity, strain and interface effect of a series of 2D perovskite oxides. The results show that the soft modes and ground-state structures of the perovskite monolayers and bilayers are significantly dependent on the perovskite tolerance factor. Ferroelectricity is absent in these perovskite monolayers while widely present in the perovskite bilayers with smaller tolerance factors. Generally, the MS ferroelectric switching path of reversing tilt mode has the lowest energy barrier. Epitaxial strain can significantly tune the amplitude of octahedral rotation distortion. Compressive strain can reduce the energy barrier of ferroelectric switching, and induce a polar to non-polar phase transition by suppressing the tilt distortion. The interface effect may significantly affect the ferroelectric switching of reversing IR mode, and even lead to the transition of the lowest-energy ferroelectric switching path.

\begin{acknowledgments}
This work was supported by the National Natural Science Foundation of China (Grant No. 11974418), the Fundamental Research Funds for the Central Universities (Grant No. 2019QNA30), the Postgraduate Research \& Practice Innovation Program of Jiangsu Province (Grant No. KYCX21\_2157), and the Assistance Program for Future Outstanding Talents of China University of Mining and Technology (Grant No. 2021WLJCRCZL170). Computer resources provided by the High Performance Computing Center of Nanjing University are greatfully acknowledged.
\end{acknowledgments}

\bibliography{reference}

\end{document}


\title{Supplemental Materials for \\Two-dimensional ferroelectricity induced by octahedral rotation distortion in perovskite oxides}
\author{Ying Zhou}
\affiliation{School of Materials Science and Physics, China University of Mining and Technology, Xuzhou 221116, China}
\author{Shuai Dong}
\affiliation{School of Physics, Southeast University, Nanjing 211189, China}
\author{Changxun Shan}
\author{Ke Ji}
\author{Junting Zhang}
\email{juntingzhang@cumt.edu.cn}
\affiliation{School of Materials Science and Physics, China University of Mining and Technology, Xuzhou 221116, China}

\maketitle

\renewcommand\thefigure{S\arabic{figure}}
\renewcommand\thetable{S\arabic{table}}

\begin{figure}
\centering
\includegraphics*[width=0.8\textwidth]{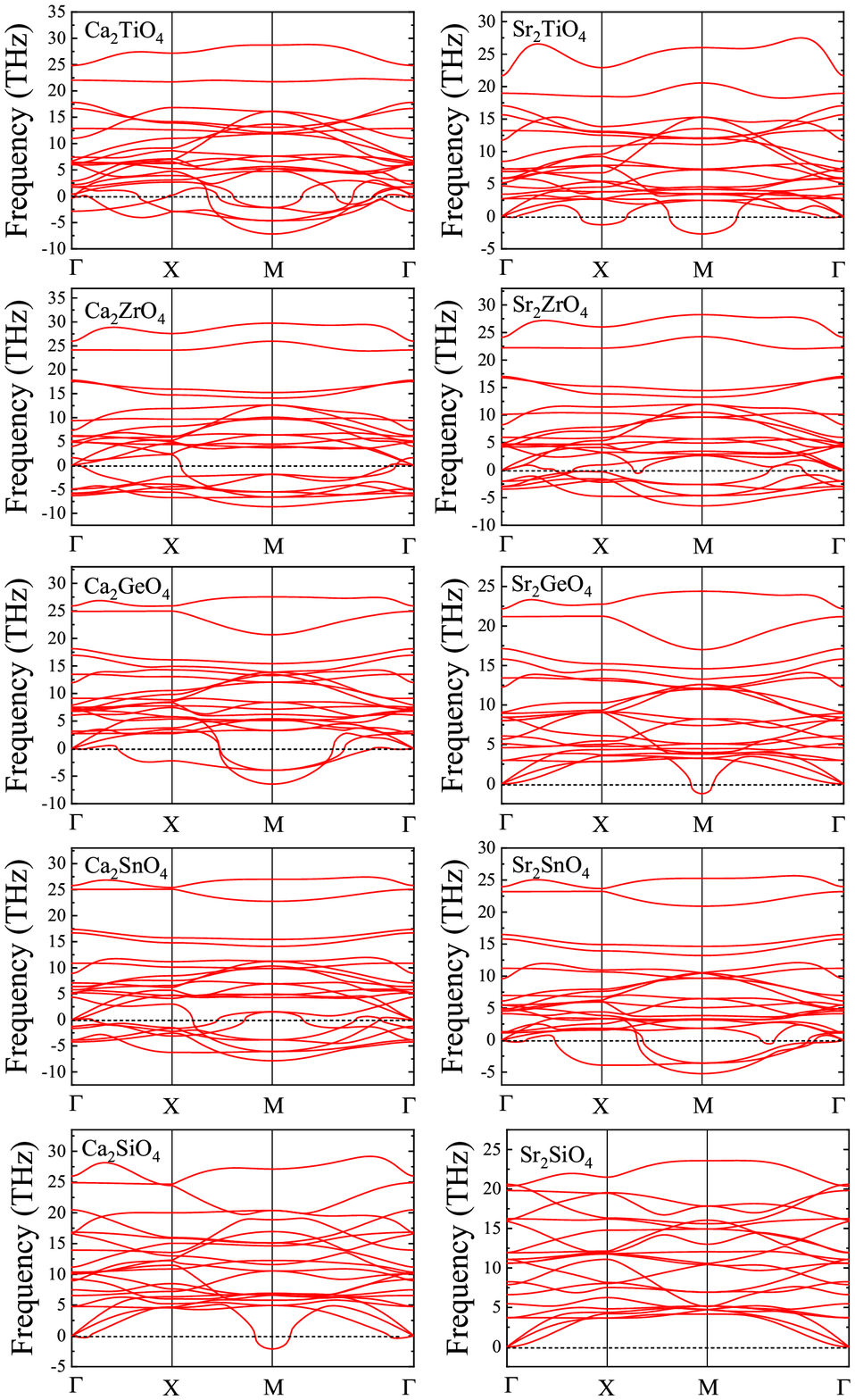}
\caption{Phonon band structures of the prototype phase ($P4/mmm$) of the perovskite monolayers.}
\end{figure}

\begin{figure}
\centering
\includegraphics*[width=0.6\textwidth]{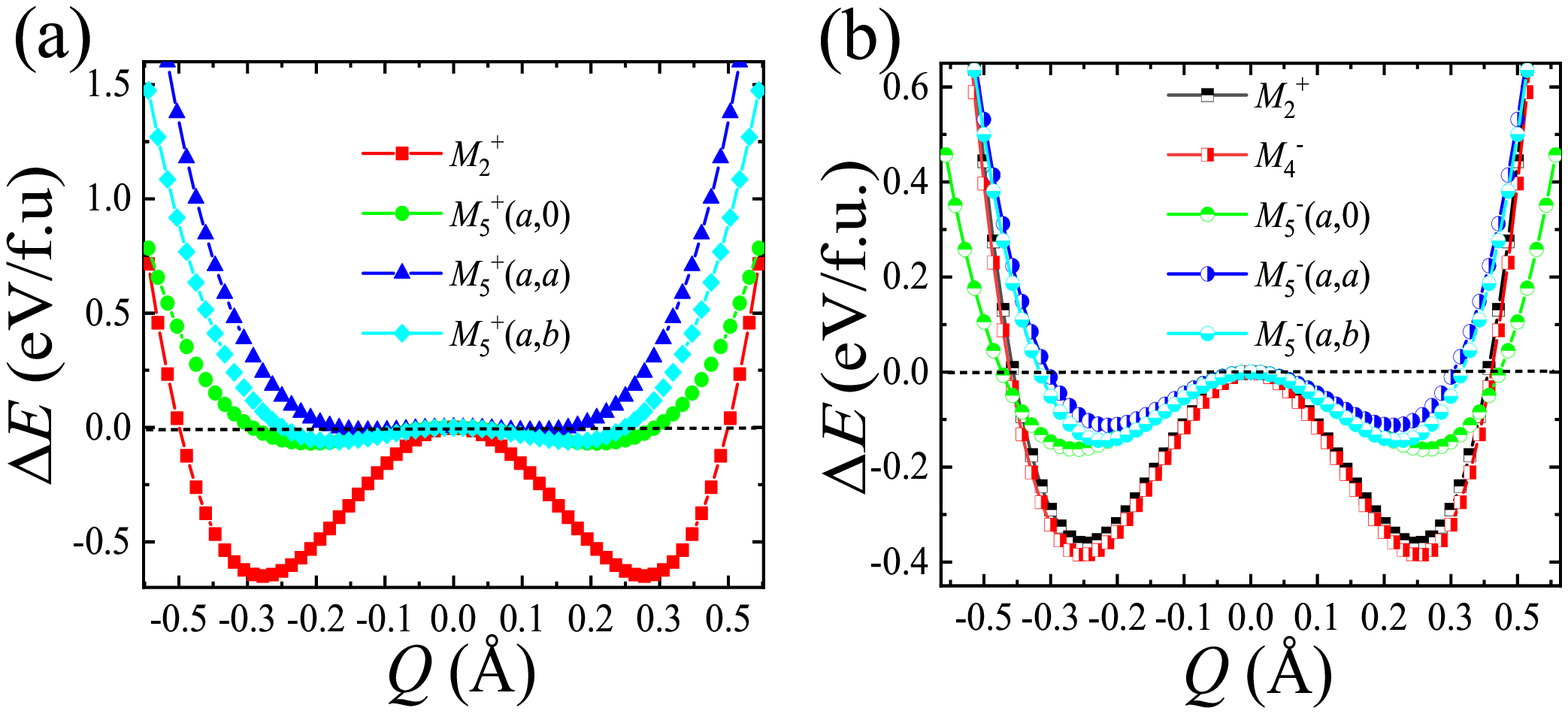}
\caption{The energy of the distorted structure as a function of the amplitude of rotation ($M_2^+$) and tilt ($M_5^+$ or $M_5^-$) modes for the perovskite (a) monolayers and (b) bilayers. The rotation mode of the perovskite bilayers is divided into in-phase ($M_2^+$) and out-of-phase ($M_4^-$) rotation. The two components of tilt mode represent the magnitude of rotation along the \emph{a} and \emph{b} axes, which are set to $b = 2a$ for $M_5^+(a,b)$ and $M_5^-(a,b)$ in the calculation.}
\end{figure}

\begin{table}
\centering
\includegraphics*[width=1\textwidth]{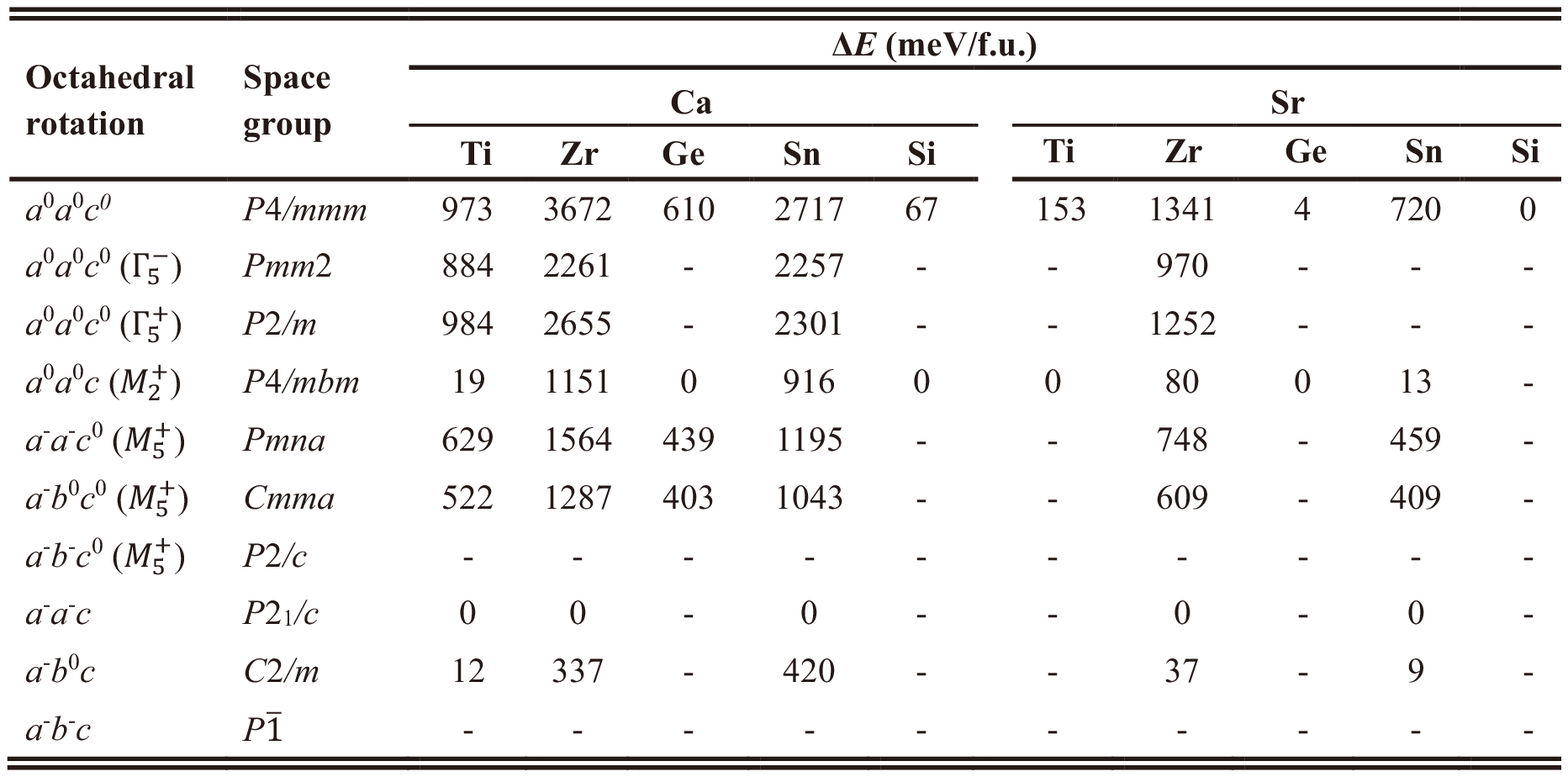}
\caption{Octahedral rotation type, symmetry and energy of each structural phase of all perovskite monolayers. The energy of each phase is given relative to that of the ground-state phase. The symbol "-" represents the absence of the corresponding structural phase, that is, the distortion mode changes or disappears after structural optimization.}
\end{table}

\begin{figure}
\centering
\includegraphics*[width=0.8\textwidth]{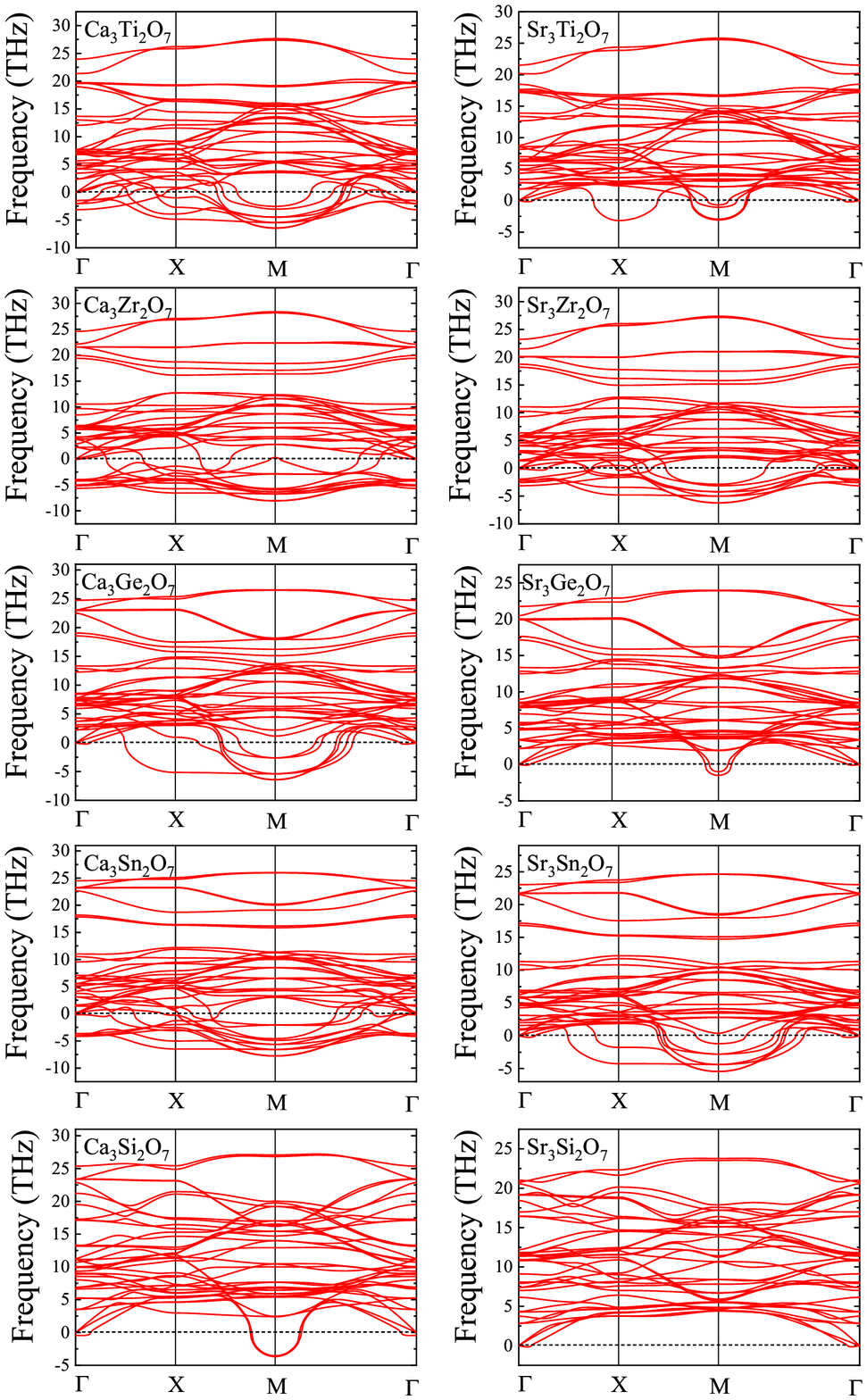}
\caption{Phonon band structures of the prototype phase of the perovskite bilayers.}
\end{figure}

\begin{table}
\centering
\includegraphics*[width=1\textwidth]{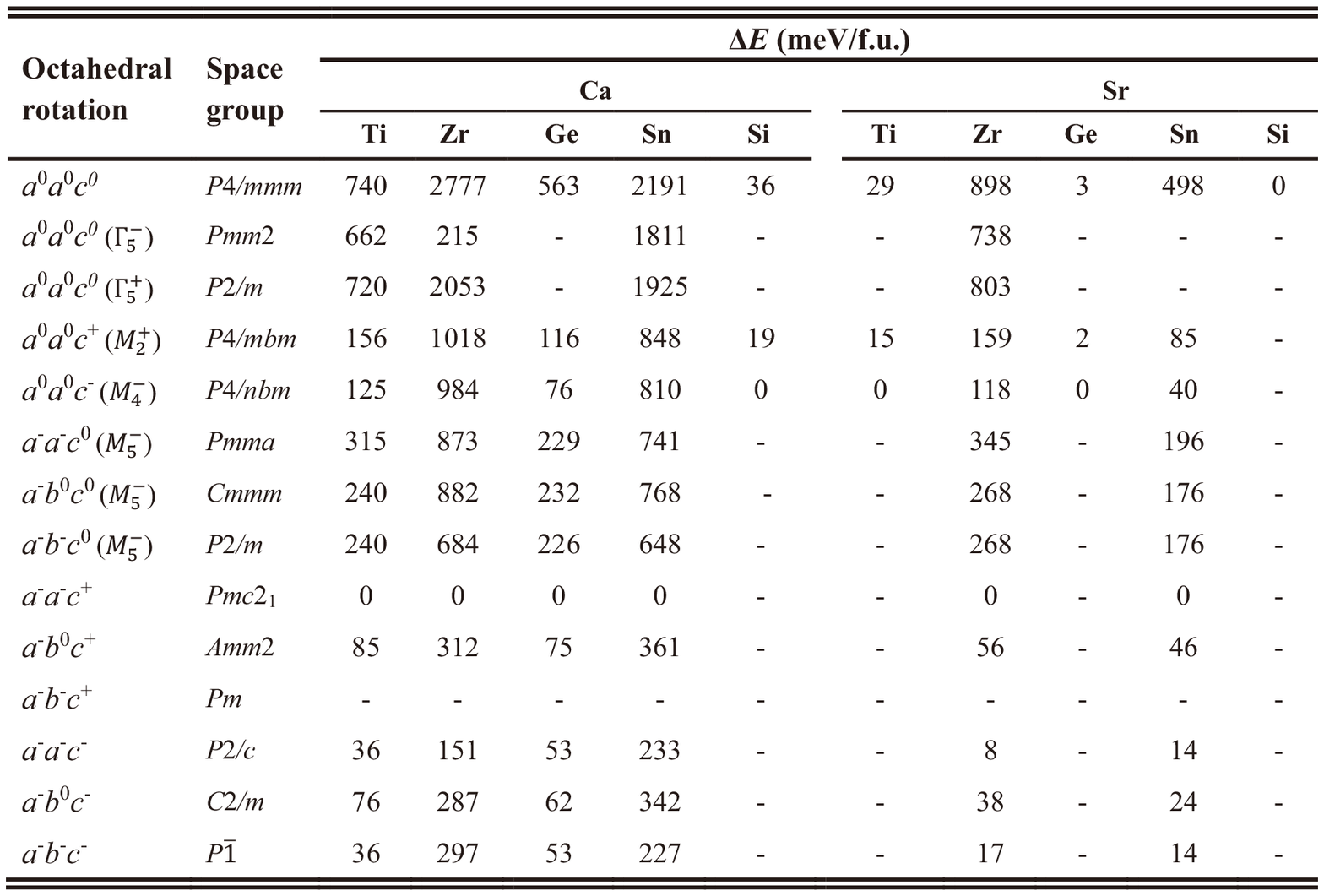}
\caption{Octahedral rotation type, symmetry and energy of each structural phase of all perovskite bilayers. The energy of each phase is given relative to that of the ground-state phase.}
\end{table}

\begin{figure}
\includegraphics*[width=0.7\textwidth]{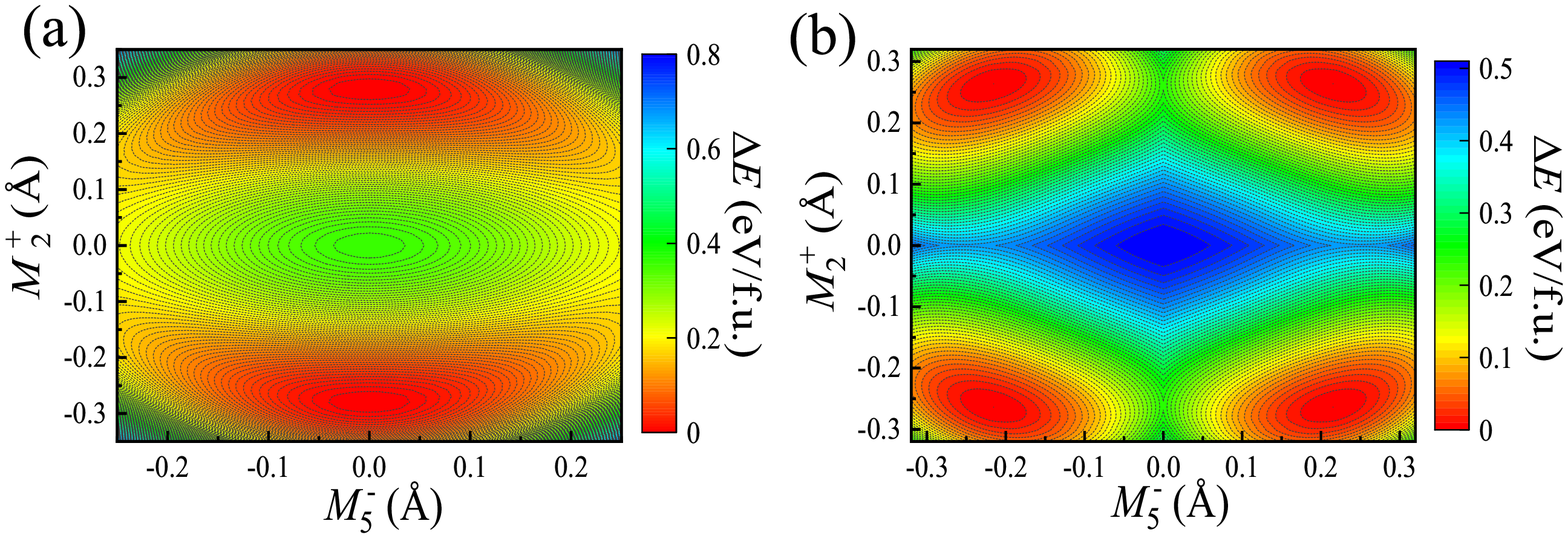}
\caption{Variation of energy with the amplitude of in-phase rotation ($M_2^+$) and tilt ($M_5^-$) modes in the (a) absence and (b) presence of the polar mode for the polar ground-state phase ($Pmc2_1$) of the Ca$_3$Ti$_2$O$_7$ bilayer.}
\end{figure}

\begin{figure}
\centering
\includegraphics*[width=0.6\textwidth]{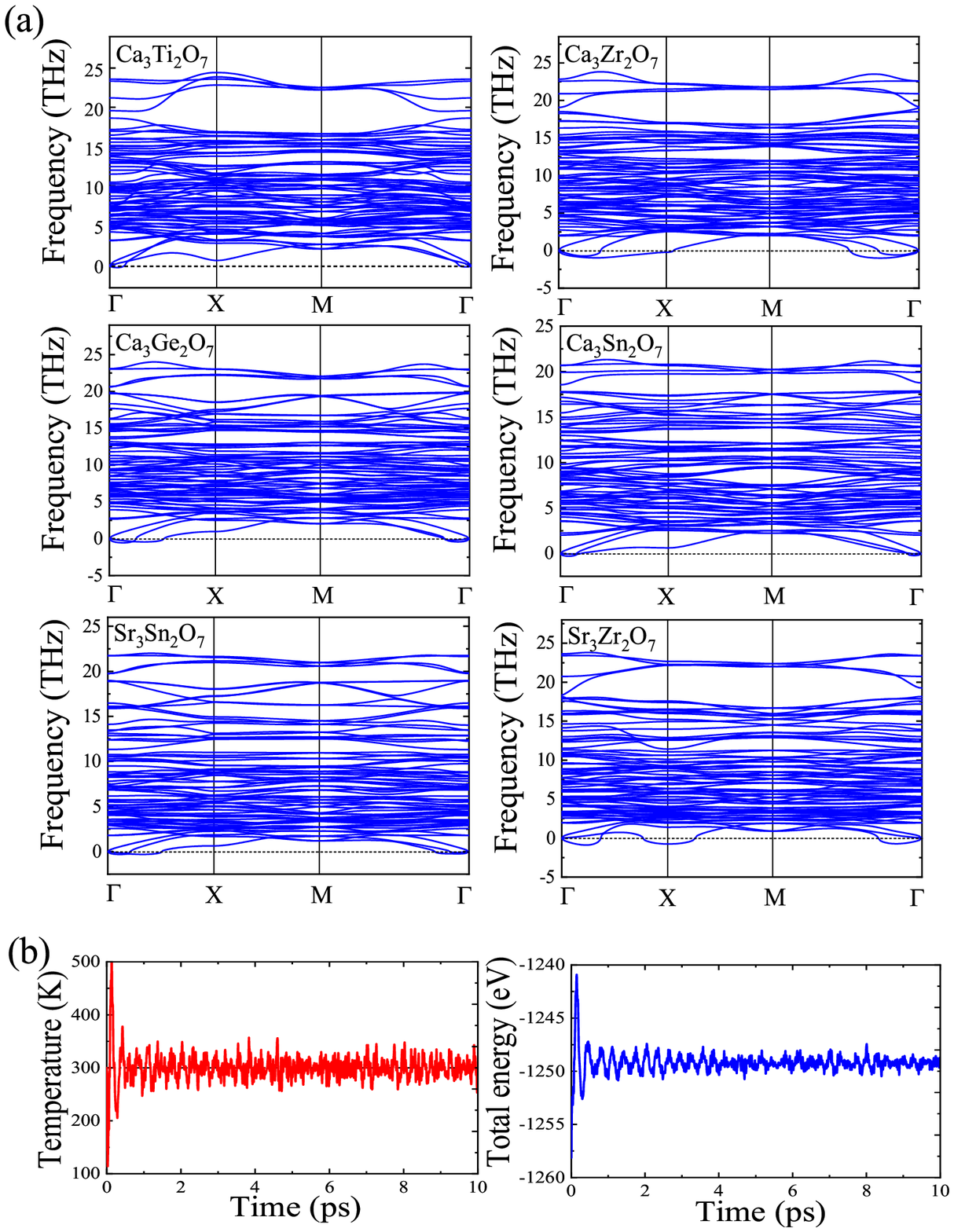}
\caption{(a) Phonon band structures of the ground-state phases of the polar perovskite bilayers. (b) First-principles molecular dynamics simulation (at 300 K) of the ground-state phase of the Ca$_3$Ti$_2$O$_7$ bilayer.}
\end{figure}

\begin{figure}
\centering
\includegraphics*[width=0.6\textwidth]{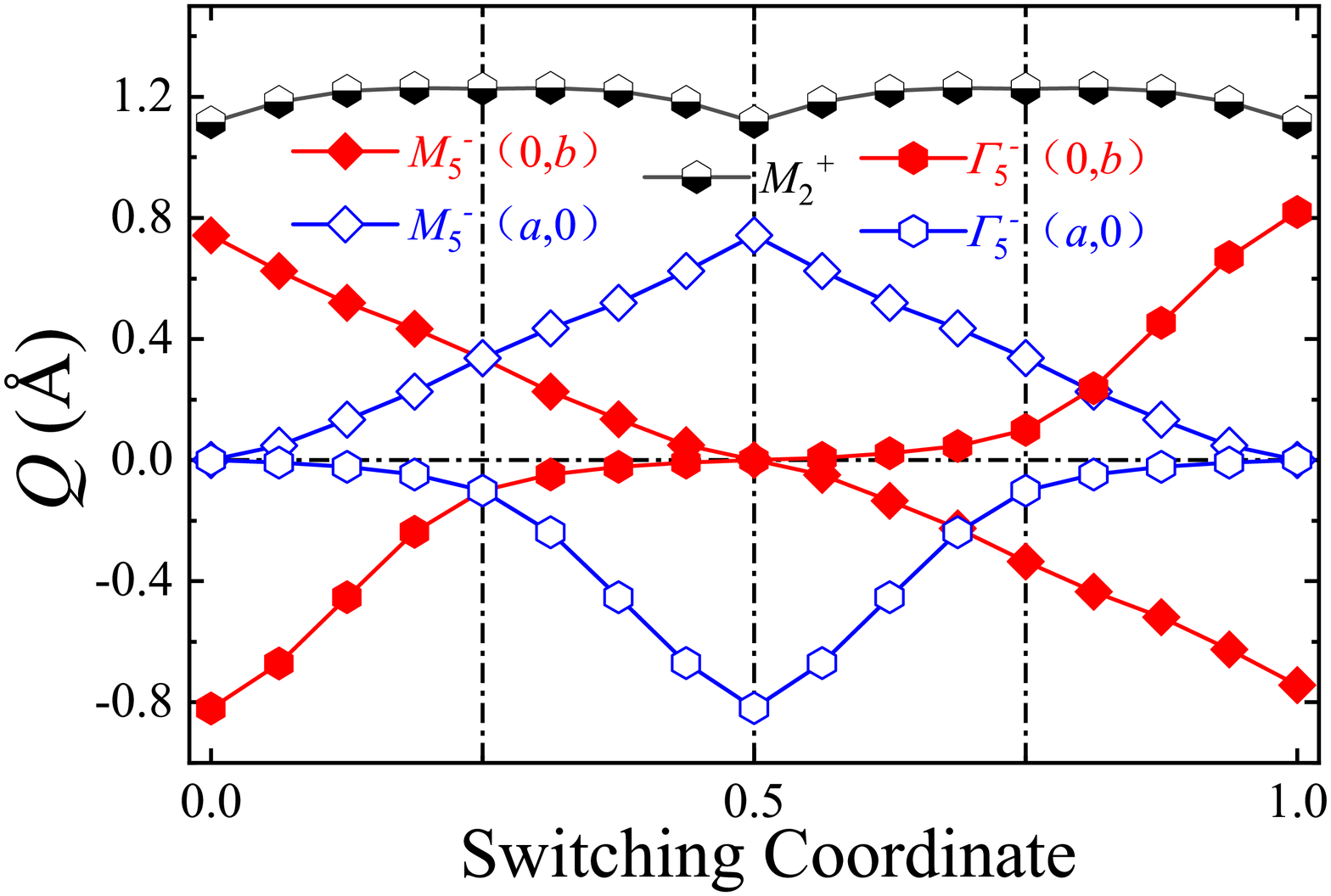}
\caption{(a) Variation in the amplitude of in-phase rotation ($M_2^+$), tilt ($M_5^-$), and polar ($\varGamma_5^-$) modes along the lowest-energy ferroelectric switching path of the Ca$_3$Ti$_2$O$_7$ bilayer.}
\end{figure}

\begin{figure}
\centering
\includegraphics*[width=1\textwidth]{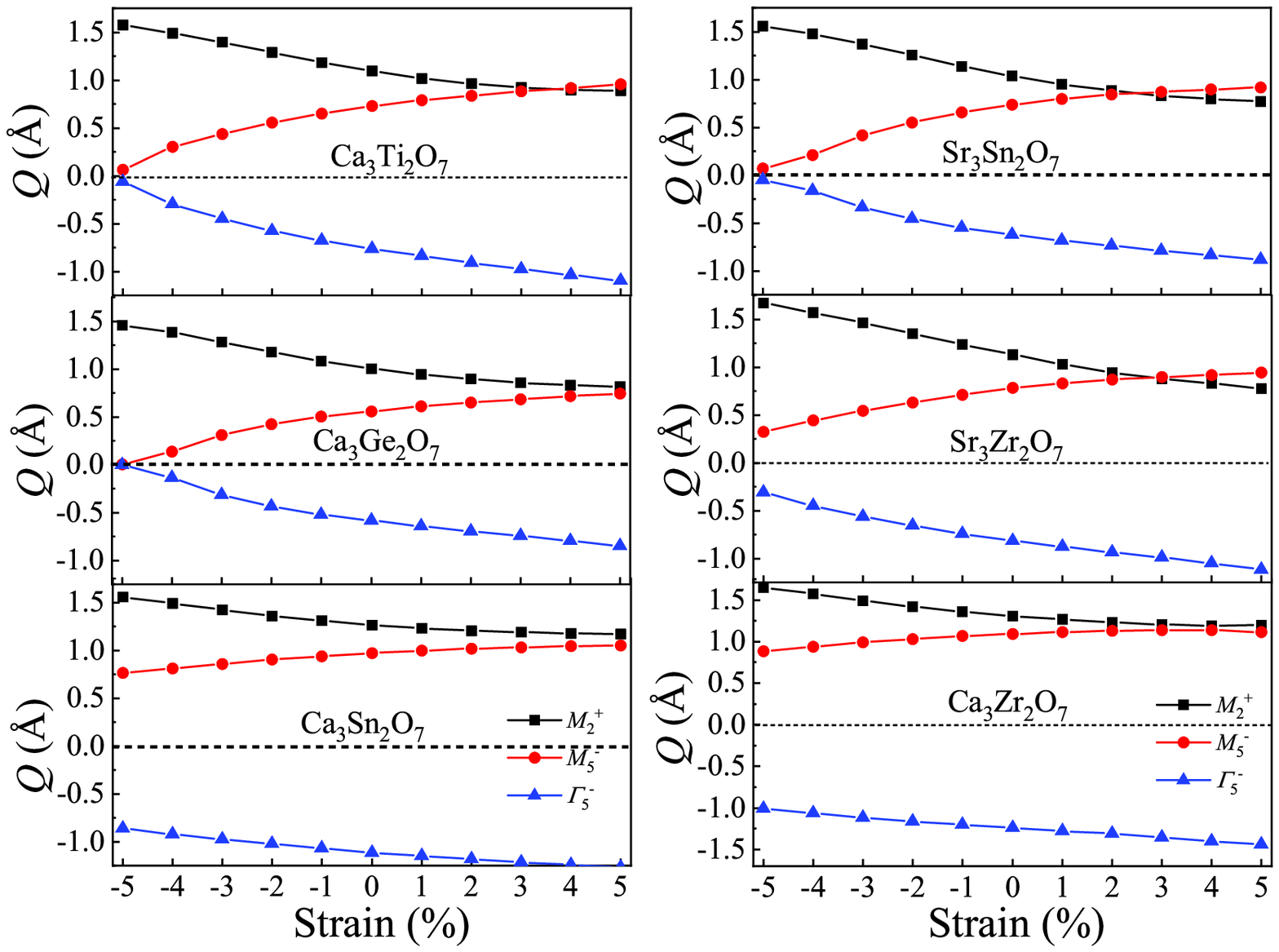}
\caption{Variation in the amplitude of in-phase rotation ($M_2^+$), tilt ($M_5^-$), and polar ($\varGamma_5^-$) modes with epitaxial strain for the polar bilayers.}
\end{figure}

\begin{figure}
\centering
\includegraphics*[width=1\textwidth]{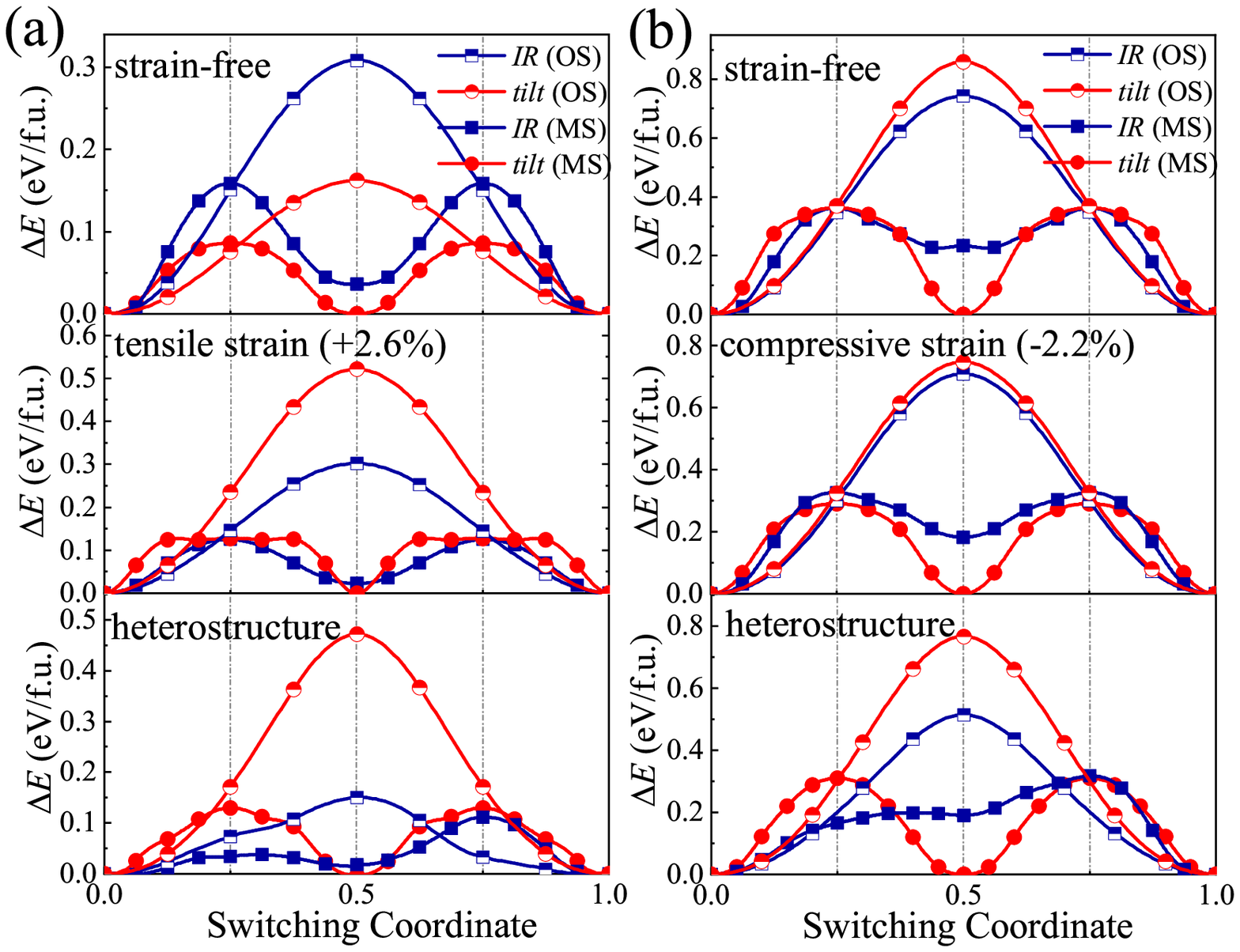}
\caption{The change of energy of the (a) Ca$_3$Ti$_2$O$_7$ and (b) Ca$_3$Sn$_2$O$_7$ bilayers as a function of switching coordinate along different ferroelectric switching paths, involving reversing the in-phase rotation (IR) and tilt modes via one step (OS) or multi steps (OS). The upper, middle and lower panels of each subfigure show the comparison between the strain-free, strained bilayers and heterostructure models. The strained bilayers are subjected to the epitaxial strain imposed by the SrTiO$_3$ substrate, therefore its strain value is equal to that of the corresponding heterostructure model.}
\end{figure}

\begin{figure}
\centering
\includegraphics*[width=1\textwidth]{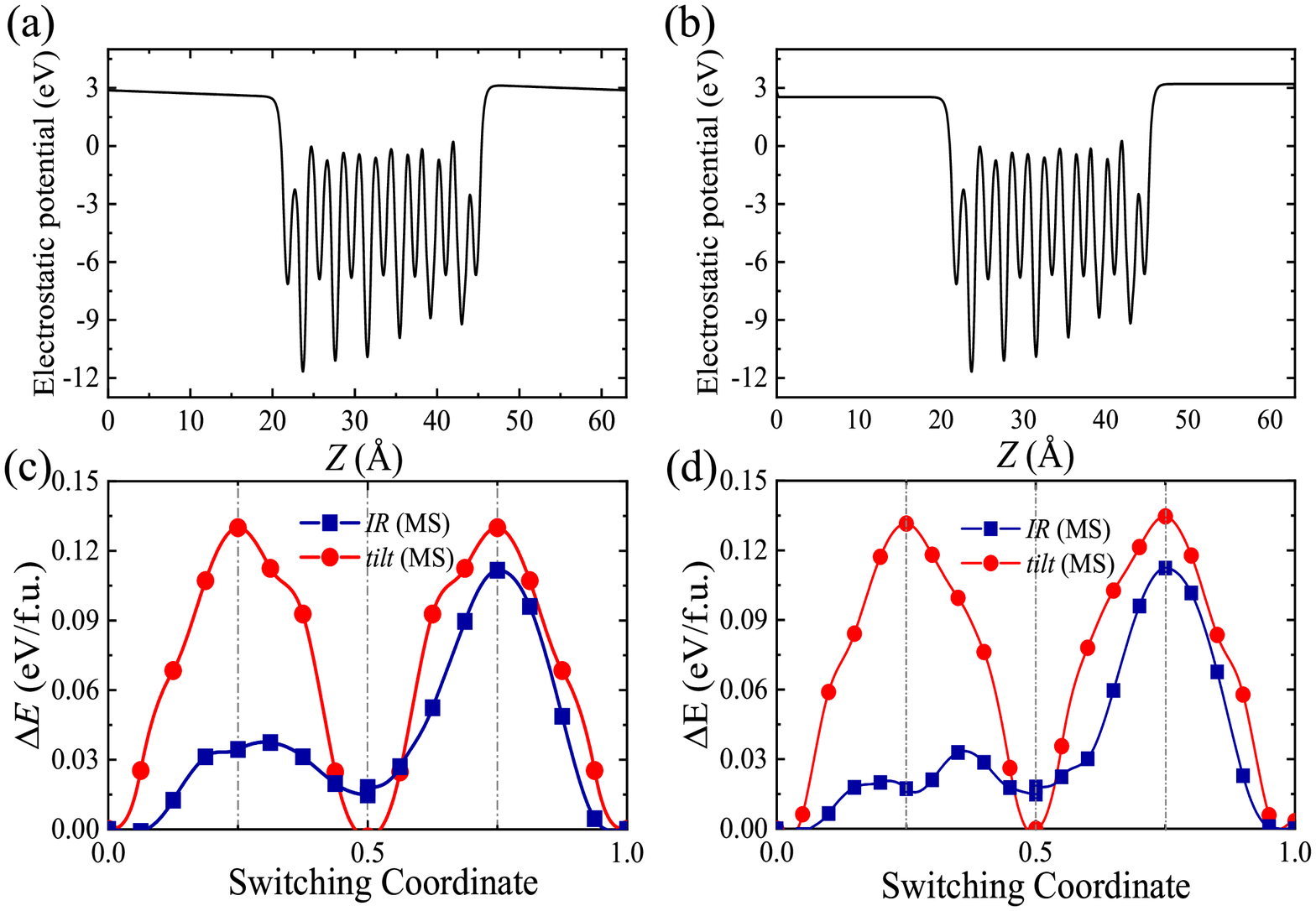}
\caption{Comparison of electrostatic potentials of the heterostructure Ca$_3$Ti$_2$O$_7$/(SrTiO$_3)_4$ calculated (a) without and (b) with dipole correction, as well as (c),(d) the corresponding energy barrier of multi-steps switching paths.}
\end{figure}